\documentclass[preprint,prd,noshowpacs,nofootinbib]{revtex4}
\usepackage{amssymb}
\usepackage{amsmath}

\begin{document}

\title{Field momentum and the reality of the Dirac String}

\author{Siva Mythili Gonuguntla}
\email{sivamythili01@mail.fresnostate.edu}
\affiliation{Physics Department, California State University Fresno, Fresno, CA 93740 USA}

\author{Douglas Singleton}
\email{dougs@mail.fresnostate.edu}
\affiliation{Physics Department, California State University Fresno, Fresno, CA 93740 USA\\}
\affiliation{Kavli Institute for Theoretical Physics, University of California Santa Barbara, Santa Barbara, CA 93106, USA}

\date{\today}

\begin{abstract}
It is well known that a magnetic monopole-electric charge system carries an angular momentum in its electromagnetic fields. Here we show that in the Dirac string formulation of magnetic charge the monopole-electric charge system also carries a momentum in its electromagnetic fields. This overlooked field momentum arises from the Coulomb electric field of the electric charge and the solenoidal magnetic field of the Dirac string. This implies that the monopole-charge system must either: (i) carry a ``hidden momentum" in the string, indicating that the string is real, or (ii) that the  monopole-charge system violates the center-of-energy theorem.  The overall conclusion is that the
Dirac monopole with a {\it regulated} Dirac string is not a true monopole, and is not even a good effective description for topological monopoles ({\it e.g.} the 't Hooft-Polyakov monopole outside the core).   
\end{abstract}

\maketitle

\section{Dirac string formulation of magnetic charge}

We briefly review the Dirac string approach to magnetic charge. By definition a magnetic charge $g$ has a Coulomb magnetic field ${\bf B} = \frac{g {\bf r}}{r^3}$. This implies that the divergence of ${\bf B}$ has a delta function source {\it i.e.}  $\nabla \cdot {\bf B} = 4 \pi g \delta ({\bf r})$. However, this last result seems to contradict the relationship between the magnetic field and the vector potential: ${\bf B} = \nabla \times {\bf A}$. If ${\bf A}$ is well behaved then $\nabla \cdot {\bf B} = \nabla \cdot (\nabla \times {\bf A}) = 0 \ne 4 \pi g \delta ({\bf r})$. 

In \cite{dirac,dirac1} Dirac proposed the following vector potentials for magnetic charge 
\begin{equation}
\label{A-coulomb}
    {\bf A} _\pm ({\bf x}) = \frac{g}{r} \left( \frac{\pm 1 - \cos \theta }{\sin \theta} \right) {\bf{\hat \varphi}} ~ =\frac{g}{\rho} \left( \pm 1 - \frac{z}{\sqrt{\rho^2+z^2}} \right) {\bf{\hat \varphi}} ~.
\end{equation}
We have given the Dirac potentials in both spherical and cylindrical coordinates for later use. Taking the curl of either ${\bf A}_\pm$ does give a Coulomb magnetic field $\nabla \times {\bf A} _\pm ({\bf x}) = \frac{g {\bf r}}{r^3}$. However, the potentials in \eqref{A-coulomb} are not well behaved in that they are singular everywhere along the $+z$-axis (for ${\bf A}_-$) or the $-z$-axis (for ${\bf A}_+$).

The standard way to handle the Dirac string singularity is by defining a regularized vector potential (see references \cite{heras,olive,blag,felsager,shnir,adorno,mavromatos} for details)
\begin{equation}
\label{A-reg}
{\bf A}^{reg}_{\pm} = \frac{g \Theta (\rho -\epsilon)}{\rho} \left( \pm 1 - \frac{z}{\sqrt{\rho^2+z^2 + \epsilon^2}} \right) {\bf{\hat \varphi}}~.
\end{equation}
 Taking the curl of \eqref{A-reg} and the limit $\epsilon \to 0$ gives
\begin{eqnarray}
\label{b-coulomb}
   {\bf B} &=&  \lim_{\epsilon \to 0} \nabla \times ({\bf A}^{reg} _\pm ) = g \frac{\rho {\hat {\bf \rho}} + z {\hat {\bf z}}}{(\rho ^2 +z^2)^{3/2}}\pm 2  g \frac{\delta (\rho)}{\rho} \Theta (\mp z) {\bf {\hat z}} \nonumber \\
   &=& g \frac{{\hat {\bf r}}}{r^2} \pm 4 \pi g \delta (x) \delta (y) \Theta (\mp z) {\hat {\bf z}} ~,
\end{eqnarray}
where we used $\frac{\delta (\rho)}{2 \pi \rho} = \delta (x) \delta (y)$. This form of the magnetic field is explicitly derived and discussed in several review articles and monographs \cite{heras,olive,blag,felsager,shnir,adorno,mavromatos} with reference \cite{heras} giving the most explicit and pedagogical derivation.
The first term on the right hand side of \eqref{b-coulomb} is the Coulomb part and the second term is the Dirac string contribution. The Coulomb part of \eqref{b-coulomb} gives an outgoing magnetic flux of $4 \pi g$ and the string part of \eqref{b-coulomb} gives an incoming flux of $4 \pi g$, thus there is a net flux of zero and $\nabla \cdot {\bf B} =0$. 

Now for a physical monopole, one only wants the Coulomb term from \eqref{b-coulomb}, thus one imposes conditions so as to make the string part of \eqref{b-coulomb} ``invisible" to any electric charge, $q$, placed in the monopole's field. There are various ways of accomplishing this (see \cite{heras,olive,blag,felsager,shnir,adorno} for details on the various approaches), but they all lead to the famous Dirac quantization condition between the magnetic and electric charge $qg= n \frac{\hbar}{2}$. The final conclusion is that the string part of \eqref{b-coulomb} has been made unphysical by imposing the Dirac quantization condition. Another argument for the unphysical character of the string part of \eqref{b-coulomb} is that the two vector potentials  ${\bf A} _+ ({\bf x})$ and ${\bf A} _- ({\bf x})$ are related by the gauge transformation ${\bf A} _+ ({\bf x})-{\bf A} _- ({\bf x}) = \nabla _{\varphi} \alpha = \frac{2g}{\rho} $with the gauge function $\alpha = 2 g \varphi$.
 
\section{Field momentum of monopole-charge system}

While the Dirac quantization condition makes the Dirac string ``invisible" from the charge, $q$, in this section we show that the string {\it is not} made invisible from the {\it electric field} of $q$. We start by placing the magnetic charge at the origin, so that the ${\bf B}$-field is given by \eqref{b-coulomb}, and we then place the electric charge at the location ${\bf r}_0$. The electric field is ${\bf E} = q \frac{{\bf\hat r}'}{{r'}^2}$ with ${\bf r}' = {\bf r} - {\bf r}_0 = (x-x_0){\bf \hat x} + (y-y_0){\bf \hat y} + (z-z_0){\bf \hat z}$. The field  momentum from the Coulomb part of the magnetic field ({\it i.e.} from $g\frac{{\hat {\bf r}}}{r^2}$) is 
\begin{equation}
\label{mom3d}
{\bf P}^{point}_{EM}= \frac{1}{4 \pi} \int  ({\bf E} \times {\bf B}) d^3 x = \frac{qg}{4 \pi} \int  \left( \frac{{\bf r}'}{{r'}^3} \times {\frac{{\bf \hat r}}{r^2}} \right) d^3 x =  0 ~.
\end{equation} 
As expected this total field momentum is zero. A non-zero field momentum for a system whose parts are not moving (the magnetic and electric charges are at rest) would have implied either: (i) There is a violation of the special relativistic center-of-energy theorem \cite{coleman,zangwill}
 or (ii) there is some hidden momentum \cite{coleman,shockley,boyer,griffiths} carried by the sources that produce the electric and magnetic fields. 
 
We now turn to the field momentum of the string part of the magnetic field in \eqref{b-coulomb} ({\it i.e.} the $\pm 4 \pi g \delta (x) \delta (y) \Theta (\mp z) {\bf {\hat z}}$ term) and show that it {\it does contribute} a non-zero part to the field momentum. It is this contribution that is the focus of this note and which we claim has been overlooked in previous work. It leads to new results for the Dirac string formulation of magnetic charge. This string contribution to the field momentum is 
\begin{eqnarray}
\label{mom3d-2a}
{\bf P}_{EM} ^{string} &=& \frac{1}{4 \pi} \int  \left (q  \frac{{\bf \hat r}'}{{r'}^2} \times ( \pm 4 \pi g \delta (x) \delta (y) \Theta (\mp z) {\bf {\hat z}}) \right) d^3 x \nonumber \\
&=& \pm gq \int _{-\infty} ^{\infty} \Theta (\mp z) \frac{x_0 {\bf \hat y} - y_0 {\bf \hat x}}{(\rho_0 ^2 + (z-z_0)^2)^{3/2}} dz \\
&=& gq \frac{r_0 \mp z_0}{r_0 \rho _0 ^2} (-y_0 {\bf \hat x} + x_0 {\bf \hat y}) \nonumber ~,
\end{eqnarray} 
where $\rho_0 ^2 = x_0 ^2 + y_0 ^2$ and $r_0 ^2 = x_0 ^2+ y_0 ^2 + z_0 ^2 = \rho_0^2 + z_0^2$. Thus this system has a non-zero field momentum ${\bf P}^{total} _{EM}= {\bf P}^{point}_{EM} + {\bf P}_{EM} ^{string} \ne 0$. Since the electric and magnetic charges are at rest, and thus have no mechanical momentum, we are left to contemplate one of the two choices listed under \eqref{mom3d}: either the center-of-energy theorem from special relativity is violated or there is some hidden momentum in the system. If we accept the first path, and assume that the center-of-energy theorem remains valid this implies that Dirac monopoles, with Dirac strings, are not allowed. Let us then look at the second path and see if this provides a resolution to this issue (we find that it does provide a resolution but at the cost of admitting the string is real). For a system to have hidden momentum there must be some physical current, and in fact there is such a current connected with the string part of the magnetic field from \eqref{b-coulomb}.    
 Inserting the magnetic field from \eqref{b-coulomb} into $\nabla \times {\bf B} = 4 \pi {\bf J}$ one immediately finds a current density 
\begin{equation}
    \label{j}
    {\bf J} = \pm g \Theta (\mp z) \left[ \delta (x) \delta ' (y) {\bf \hat x} - \delta (y) \delta' (x) {\bf \hat y} \right] ~,
\end{equation}
where primes indicate differentiation with respect to the argument of the delta function. 
This current density comes only from the string part of \eqref{b-coulomb} since the Coulomb part of the magnetic field has zero curl. We have worked out ${\bf J}$ in Cartesian coordinates since this will simplify subsequent calculations. 

It is possible that the current density in \eqref{j} plus the charge $q$ could carry ``hidden" mechanical momentum which could balance the electromagnetic field momentum in \eqref{mom3d-2a} thus saving the center-of energy theorem. To check this take the charge to be located at ${\bf r}_0 = (x_0, 0, z_0)$ {\it i.e.} $y_0=0$. This can be done without loss of generality due to the cylindrical symmetry of the system with respect to the string directions. With this location for the charge $q$ equation \eqref{mom3d-2a} gives 
\begin{equation}
\label{mom3d-2b}
{\bf P}_{EM} ^{string} = gq \frac{r_0 \mp z_0}{r_0 x_0}{\bf \hat y} ~,
\end{equation}
where $r_0 =\sqrt{x_0^2 + z_0 ^2}$ in this case. The hidden mechanical momentum contained in the sources $q$
 and ${\bf J}$ is 
 \begin{equation}
     \label{hid-mom}
     {\bf P}_{mech} ^{hid} = - \int \phi {\bf J} d^3x
 \end{equation}
 Reference \cite{griffiths} gives a good discussion of hidden mechanical momentum, along with a derivation of \eqref{hid-mom}.
 For the set up with $q$ at ${\bf r}_0 = (x_0, 0, z_0)$ the potential is $\phi =\frac{q}{\sqrt{(x-x_0)^2 +y^2+(z-z_0)^2}}$. We now specialize to the case when the string is along the $-z$ axis (the case with the string along the $+z$ axis works out similarly except for a few changes of sign) so that ${\bf J} = + g \Theta (- z) \left[ \delta (x) \delta ' (y) {\bf \hat x} - \delta (y) \delta' (x) {\bf \hat y} \right]$ equation \eqref{hid-mom} becomes
 \begin{equation}
     \label{hid-mom-a}
     {\bf P}_{mech} ^{hid} = - gq \int_{-\infty} ^0 dz \int _{-\infty} ^{\infty} dy \int _{-\infty} ^{\infty} dx
     \frac{\left[ \delta (x) \delta ' (y) {\bf \hat x} - \delta (y) \delta' (x) {\bf \hat y} \right]}{\sqrt{(x-x_0)^2 +y^2+(z-z_0)^2}}  ~.
 \end{equation}
The $\Theta (-z)$ function has been taken into account via the limits on the $z$-integration. To handle the $\delta '(y)$ and $\delta ' (x)$ terms one uses integration by parts. For the $dy$ integration, of the ${\bf \hat x}$ term, integration by parts gives
$$
\frac{\delta (y)}{\sqrt{(x-x_0)^2 +y^2+z^2}} {\bigg \vert}_{-\infty} ^{~\infty} + \int _{-\infty} ^{\infty} \frac{y \delta (y)}{((x-x_0)^2 +y^2+(z-z_0)^2)^{3/2}} dy~.
$$
The first term is zero since the function vanishes at $\pm \infty$ and the second term is zero due to the $y \delta (y)$ term in the integrand. Thus there is no hidden mechanical momentum in the ${\bf \hat x}$ direction which is consistent with the result of \eqref{mom3d-2b}.
 
Next we carry out the inetgration of the second term in \eqref{hid-mom-a}. The $dy$ integration is trivial and simple sets $y=0$ in the rest of the integrand. The $dx$ integration is handled via integration by parts and yields
$$
\frac{\delta (x)}{\sqrt{(x-x_0)^2 +(z-z_0)^2}} {\bigg \vert}_{-\infty} ^{~\infty} + \int _{-\infty} ^{\infty} \frac{(x-x_0) \delta (x)}{((x-x_0)^2 +(z-z_0)^2)^{3/2}} dx~.
$$
The first, surface term is zero since the function vanishes at $\pm \infty$ and, after doing the $dx$ integration, the second term becomes  $-\frac{x_0}{(x_0^2 + (z-z_0)^2)^{3/2}}$. Collecting together all the integrations and various minus signs, \eqref{hid-mom-a} becomes
 \begin{eqnarray}
     \label{hid-mom-b}
     {\bf P}_{mech} ^{hid} &=& - gq x_0 \int_{-\infty} ^0 dz 
     \frac{{\bf \hat y}}{(x_0^2 + (z-z_0)^2)^{3/2}} = - gq x_0 \left( \frac{1}{x_0 ^2} - \frac{z_0}{x_0^2 \sqrt{x_0^2 + z_0^2}}\right){\bf \hat y} \nonumber \\
     &=& -gq\frac{(r_0 - z_0)}{r_0 x_0} {\bf \hat y}~.
 \end{eqnarray}
 This hidden mechanical momentum balances the electromagnetic field momentum from \eqref{mom3d-2b} for the case with the string along the $-z$-axis ({\it i.e.} ${\bf P}_{EM} ^{string}+{\bf P}_{mech} ^{hid}=0$). This restores the center-of-energy theorem, {\it but} now one has to allow that the string is a real physical entity since it must carry mechanical momentum \eqref{hid-mom-b} to balance the field momentum \eqref{mom3d-2b}. In the conclusion we will argue that this latter option - that the Dirac monopole, with a regularized string, does not represent a real monopole, but rather a monopole-like field at the end of a real, but infinite solenoid.  
 
\section{Conclusions}

In this short work we showed that there is a heretofore overlooked contribution to the electromagnetic field momentum of a Dirac monopole with a Dirac string plus an electric charge. This field momentum came from the interaction of the magnetic field of the Dirac string and the electric field of the charge. Hiding the magnetic field of the Dirac string from the charge $q$ is how one arrives at the famous Dirac quantization condition, $qg = n \frac{\hbar}{2}$. There are a host of different ways of hiding the Dirac string from the charge $q$ to obtain the Dirac quantization condition \cite{heras,olive,blag,felsager,shnir,adorno,jackson}. However, in this work we find that one can not hide the magnetic field of the Dirac string from {\it the electric field} of $q$. The electric field of $q$ and the magnetic field of the Dirac string (the second term in \eqref{b-coulomb}) produce a non-zero field momentum given by \eqref{mom3d-2a}. One might have expected that since the Dirac string is infinitesimally thin, that this would lead to the vanishing of the field momentum, but the simple, direct calculation of \eqref{mom3d-2a} shows this not to be the case - there is a non-zero field momentum. Thus one has the two previously mentioned options: (i) either ${\bf P}^{string} _{EM}$ is non-zero and unbalanced which violates the center-of-energy theorem \cite{coleman,zangwill} or (ii) there is a hidden mechanical momentum in the sources $q$ and ${\bf J}$ associated with the electric charge and magnetic field of the Dirac string. If one chooses option (i) then either one must conclude Dirac monopoles with Dirac strings are not viable models of magnetic charge or one must accept the violation of the center-of-energy theorem. This is a terrible option since violating the center-of-energy theorem  is an extremely unpalatable choice. If one chooses option (ii) then the center-of-energy theorem is saved since there is a hidden mechanical momentum to balance the field momentum. However, now that string has become a real, physical entity since it must carry mechanical momentum.
In our view option (ii) is the correct one. It points to the fact that the Dirac monopole construction with a regularized string (see equations \eqref{A-reg} and \eqref{b-coulomb}) is not a
monopole, but is simply a standard configuration of electric charges which generates a monopole-like magnetic field plus a solenoidal magnetic flux. 

One might hope that the Wu-Yang fiber bundle approach \cite{wu-yang} to magnetic charge, which avoids the Dirac string would offer a consistent model for magnetic charge. However, one can also show that this approach has a previously overlooked field momentum \cite{sps} very similar to that found in this work. This field momentum in the Wu-Yang approach also spoils this model for magnetic charge.     

The above analysis does not have anything to say about non-Abelian \cite{thooft,polyakov} monopoles or recently discovered electroweak monopoles \cite{hung1,hung2}, which owe their existence to non-trivial topology as opposed to a Dirac string. As well this analysis does not rule out the two vector potential \cite{cabibbo} or photon-dual photon approach to magnetic charge \cite{singleton-95,singleton-98}.

In a recent work \cite{mpd} it was also found that the Dirac string magnetic field also contributed a previously overlooked contribution to the well known electromagnetic field angular momentum \cite{heras,olive,blag,felsager,shnir,adorno,mavromatos,jackson} to a charge-monopole system. However, while the field angular momentum coming from the magnetic field of the string plus electric field of the charge, did significantly alter the analysis of the total angular momentum of the system, it was found \cite{mpd} that it was still possible to give a consistent description of the total field angular momentum. \\

{\bf Acknowledgment:} DS is supported by a 2023-2024 KITP Fellows Award. This research was supported in part by the National Science Foundation under Grant No. NSF PHY-1748958. \\


\begin{thebibliography}{99}

\bibitem{dirac} P.A.M. Dirac, 
Proc. Roy. Soc. A {\bf 133}, 60-72 (1931).

\bibitem{dirac1} P.A.M. Dirac, 
Phys. Rev. {\bf 74}, 817-830 (1948).

\bibitem{heras} R. Heras, Contemp. Phys. {\bf 59}, 331 (2018).

\bibitem{olive} P. Goddard and D. I. Olive, Rep. Prog. Phys. {\bf 41}, 1357 (1978).

\bibitem{blag} M. Blagojevi{\'c} and P. Senjanovi{\'c}, Phys. Rept., {\bf 157}, 233 (1988)

\bibitem{felsager} B. Felsager, {\it Geometry, Particles, and Fields}, (Springer-Verlag, New York, 1998). 

\bibitem{shnir} Y.M. Shnir, {\it Magnetic Monopoles} (Springer, Berlin 2005).

\bibitem{adorno} T.C. Adorno, D.M. Gitman, and A.E. Shabad, Proc. of the Steklov Inst. of Math. {\bf 309}, 1 (2020).

\bibitem{mavromatos} N. E. Mavromatos and V. A. Mitsou, Int. J. Mod. Phys.A {\bf 35}, 2030012 (2020).

\bibitem{coleman} S. Coleman and J. H. Van Vleck, 
Phys. Rev. {\bf 171}, 1370 (1968).

\bibitem{zangwill} A. Zangwill, {\it Modern Electrodynamics}, (Cambridge University Press, Cambridge, UK 2013) pgs. 519-522.

\bibitem{shockley} W. Shockley and R. P. James, Phys. Rev. Lett. {\bf 18}, 876 (1967).

\bibitem{boyer} T. Boyer, Am. J. Phys. {\bf 73}, 953 (2005).

\bibitem{griffiths} D. Babson, S. P. Reynolds, R. Bjorkquist and D. J. Griffiths, 
Am. J. Phys. {\bf 77}, 826 (2009). 

\bibitem{jackson} J.D. Jackson, {\em Classical Electrodynamics}  $2^{nd}$ edition, (John Wiley \& Sons, 1975).

\bibitem{wu-yang} T.T. Wu and C.N. Yang, Phys. Rev. D {\bf 12}, 3845 (1975).

\bibitem{sps} S. Gonuguntla and D. Singleton, ``Revisiting the Wu-Yang approach to magnetic charge", e-Print: 2308.14706 [hep-th]

\bibitem{thooft} G. 't Hooft, Nucl.Phys.B {\bf 79}, 276 (1974).

\bibitem{polyakov} A. M. Polyakov, JETP Lett. {\bf 20}, 194 (1974).

\bibitem{hung1} P.Q. Hung, Nucl. Phys.B {\bf 962}, 115278 (2021).

\bibitem{hung2} J. Ellis, P.Q. Hung, and Nick E. Mavromatos, Nucl. Phys.B {\bf 969}, 115468 (2021).

\bibitem{cabibbo} N. Cabibbo and E. Ferrari, Nuovo Cim. {\bf 23}, 1147 (1962).

\bibitem{singleton-95} D. Singleton, Int. J. Theor. Phys. {\bf 34}, 37 (1995).

\bibitem{singleton-98} D. Singleton, 
Am. J. Phys. {\bf 66}, 697 (1998).

\bibitem{mpd} M. Dunia, P.Q. Hung, and D. Singleton, Eur. Phys. J. C {\bf 83}, 487 (2023). 







\end{thebibliography}
\end{document}